# Does consciousness really collapse the wave function?:

# A possible objective biophysical resolution of the measurement problem

Fred H. Thaheld*
99 Cable Circle #20
Folsom, Calif. 95630 USA

**Abstract**

An analysis has been performed of the theories and postulates advanced by von Neumann, London and Bauer, and Wigner, concerning the role that consciousness might play in the collapse of the wave function, which has become known as the measurement problem. This reveals that an error may have been made by them in the area of biology and its interface with quantum mechanics when they called for the reduction of any superposition states in the brain through the mind or consciousness. Many years later Wigner changed his mind to reflect a simpler and more realistic objective position, expanded upon by Shimony, which appears to offer a way to resolve this issue. The argument is therefore made that the wave function of any superposed photon state or states is always objectively changed within the complex architecture of the eye in a continuous linear process *initially* for most of the superposed photons, followed by a discontinuous nonlinear collapse process *later* for any remaining superposed photons, thereby guaranteeing that only final, measured information is presented to the brain, mind or consciousness. An experiment to be conducted in the near future may enable us to simultaneously resolve the measurement problem and also determine if the linear nature of quantum mechanics is violated by the perceptual process.

* e-mail address: *fthaheld@directcon.net*

# 1. Introduction

What is the measurement problem and why is the act of measurement deemed so important in quantum mechanics that it has engendered such spirited discussion for over 7 decades? The measurement problem can be approached in the following fashion. von Neumann (1932) advanced the theory that the possible states of a system can be characterized by state vectors, also known as wave functions, which change in two ways: continuously in a linear fashion as a result of a passage of time, as per Schrödinger's equation and, discontinuously if a measurement is carried out on the system (Wigner, 1961; Shimony, 1963). This second type of discontinuous change, called the reduction of the state vector or collapse of the wave function, is unacceptable to many physicists. The measurement problem can then be posed as how and when does the wave function collapse or, how does a state reduction to one of the eigenstates of the measured observable occur.

von Neumann (1932) was the first one to conceive this problem in terms of what is known as the 'von Neumann chain' (Esfeld, 1999). He starts with a quantum object, an observable of which is to be measured. However, based on the formalism of quantum theory and the Schrödinger dynamics in particular, as a result of the interaction between the object and the measuring instrument, the object is entangled with the instrument. von Neumann extends this chain up to an observer but, if we take an observer into consideration, we simply end up with a description according to which the body of the observer, including his or her brain, is entangled with the instrument and the object. The measurement problem can then be further refined as to how it is that a state reduction to one of the eigenstates of the measured observable can occur in this chain (Esfeld, 1999).

von Neumann showed that as far as final results are concerned, you can cut the chain and insert a collapse anywhere you please (Herbert, 1985). He felt that the process by which a physical signal in the brain becomes an experience in the human mind or human consciousness, is the site of the wave function collapse.

London and Bauer (1939) have postulated that consciousness randomly selects one product state out of the superposition of product states, thereby effecting a state reduction. Wigner (1961) feels that consciousness or the mind, plays a more directly physical role, adding an extra term to the mathematical equations and hence, selecting one particular branch of the wave function and one particular result for the experiment, thus producing the effect that von Neumann called collapse. They feel therefore, that this is a subjective rather than an objective process. Like von Neumann and, London and Bauer before him, Wigner did not give many details to back up his idea (Whitaker, 1996, p. 201; Esfeld, 1999). Wigner readily concedes that we do not have at our disposal a description of how a state reduction is effected by consciousness. He goes on further to state, with remarkable candor that, "We are facing here the perennial question, whether we physicists do not go beyond our competence when searching for philosophical truth. I believe that we probably do" (Wigner, 1963).

He suggests that the dynamics of quantum theory has to be modified in such a way that events of state reduction by consciousness are taken into account (Wigner, 1963). And, he further feels that the equations of motion of quantum mechanics cease to be linear, in fact they are grossly nonlinear if conscious beings enter the picture.

Shimony (1963) feels that the conceptual problems of quantum mechanics will be resolved by discovering corrections to the physical theory itself, for example, by finding that the time-dependent Schrödinger equation is only an approximation to an exact nonlinear equation governing the evolution of the state of a system. If this proves to be true, then he advances the theory that the reduction of a superposition could perhaps occur when the microscopic system interacts with the macroscopic apparatus, and no appeal to the consciousness of an observer for this purpose would be required.

In his later years Wigner changed his position to an exactly opposite viewpoint, in order to avoid solipsism and the role that consciousness plays, closely mirroring that postulated by Shimony (Mehra, 1995, p. 593). Wigner considered it to be necessary to admit state reductions independently of an observer's consciousness, and makes a concrete suggestion for an amendment of the Schrödinger equation which is intended to describe a physical process of state reduction (Mehra, 1995. pp. 73, 230). A state reduction is now felt by Wigner to occur as an *objective* event in the physical realm before the von Neumann chain reaches the consciousness of an observer (Mehra, 1995, pp. 75-77; 242-243; Esfeld, 1999).

It is of historic interest to note here that Dirac (1930), who invented the idea of wave function collapse, said that it is *nature* that makes the choice of measuremental result; once made the choice is 'irreversible and will affect the entire future state of the world' (Kragh, 1990).

It is felt by the author that Shimony and Wigner are pointing us in a direction which may finally lead to a resolution of the measurement problem through a cooperative blending of certain philosophical, theoretical and empirical aspects in the realms of quantum mechanics and biology, with emphasis on the eye in the role of a *living* macroscopic measuring instrument, serving to terminate the von Neumann chain in an objective discontinuous nonlinear fashion.

**2. The eye as a macroscopic apparatus and its possible role in resolving the measurement problem**

One can more readily grasp the essentials of the measurement problem from a biological viewpoint by studying the diagram of the visual pathways in primates in Fig. 1 (Dowling, 1987). Visual information from the eyes in the form of retinal ganglion cell spike trains passes, via the optic nerve, to the lateral geniculate nucleus and is relayed from there via the optic radiation to the visual cortex. In primates the eyes face forward and the visual fields of each eye overlap. Information from the right visual field in the form of photons is received by the left half of each retina and vice versa. At the optic chiasm, the fibers from the right and left sides of the two retinas are sorted out so that information from the right visual field projects to the left side of the brain and information from the left visual field projects to the right side of the brain.

Further compounding this problem is that the eye, which is a most complex piece of apparatus (Fig. 2), seems to have been treated in a very simplistic fashion with regards to the measurement problem, with only the retina being considered as possibly playing a role in the measurement problem and, a limited one at that. I have pointed out in previous papers (Thaheld, 2000; 2003) how the eye mistakenly got relegated to a minor, almost nonexistent role, when it should have been accorded a more prominent position. Without belaboring the point again, several physicists repeatedly cited experimental results (or each other's references!) incorrectly, showing that the threshold for visual perception in humans varies from 3-7 photons or quanta impinging upon the retina, without taking into consideration how many more photons are required initially to achieve this threshold due to the high losses sustained in their passage (Hecht et al, 1942).

What von Neumann, London and Bauer, and Wigner (initially) are claiming is that superposed states which are received in the form of photons by the eyes (Fig. 1,2) and are transduced or converted into analogous electrical information, move on as shown in the diagram until they are received by the primary visual cortex and the brain. A state reduction is supposedly then brought about only when the consciousness of an observer is reached (Wigner, 1961; 1963). If one examines this matter in detail, it begins to appear that they were incorrect in their assumptions, based on the following analysis.

There appear at first glance to be many different regions within the eye which could cause a linear change of the wave function *initially,* followed *later* and *sequentially* by a discontinuous nonlinear collapse taking place each time in the same specific region. And, that redundancy almost seems to be built in by nature for a reason. Any incident photons have to run a very daunting gauntlet before they are even converted or transduced to retinal ganglion cell spike trains. Referring to Fig. 2, there is a 4% loss as a result of corneal reflection, 50% losses due to ocular media absorption (involving both the aqueous and vitreous humor), and 80% retinal transmission losses (Hecht et al, 1942). The quantum detection efficiencies of the retinal rods (which is itself a stochastic process) range from 25% to 36%, and falls within the 80% retinal transmission losses. To give you some idea of the complex biochemistry to which the photons are initially subjected prior to arriving at the retina (Whikehart, 2003), the aqueous humor contains in its fluid, albumin, ascorbate, bicarbonate, calcium, cholesterol, globulin, glucose, hydrogen ions (as pH), phosphate, potassium, sodium and triacylglycerols.  The vitreous humor consists of approximately 40% gel and 60% fluid, containing ascorbate, bicarbonate, glucose, hyalurin, potassium, protein and sodium.  At this point, prior to impinging upon the retina, there has already been a loss of 55% of the photons which were initially incident upon the cornea!

You can get a better feeling for some of the barriers faced by the remaining photons after arriving at the retina (which is only 200-250 µm thick), and the further complexity involved, by referring to the schematic diagram of the vertebrate retina in Fig. 3 (Meister et al, 1994; Rieke, Baylor, 1998) and, my paper outlining a measurement problem experiment utilizing retinal tissue (Thaheld, 2003).  This shows the 5 major types of neurons contained in the retina, with the rod and cone photoreceptors (P) connected to bipolar cells (B), as well as horizontal cells (H).  Bipolar cells in turn make synapses with amacrine cells (A) and retinal ganglion cells (G), whose axons (Ax) form the optic nerve.  Photoreceptors send electrochemical signals to the brain by both direct (cone) and indirect (cone and rod) synaptic mechanisms.  In addition, there exist very sophisticated modulation systems that are facilitated by the horizontal and amacrine cells, among others.  As a further example of this complexity, the dendrites of ganglion cells reach up into the inner retina and read out activity formed by interactions of bipolar and amacrine cells.  Also, signal propagation between synapses can occur by physical contact with a bi-directional flow of ions, as in the ganglion cells and most neurons or, by chemicals known as neurotransmitters.

When one examines even more complex schematics involving the anatomical, physiological and neurochemical aspects of the retina (Whikehart, 2003, pp. 239-246), one can begin to envision that you are looking at a very efficient living, but not conscious, macroscopic measuring apparatus which would pose a formidable barrier to any superposed states.

When I first saw these diagrams years ago (Meister et al, 1994; Rieke, Baylor, 1998) and, not being conversant with the eye, I immediately thought that a mistake had been made either by the authors or by nature herself and, that this whole arrangement seemed to be turned around, in that light must pass through the entire thickness of the retina before striking the photoreceptors! Why would one force the photons to bypass these 5 neurons, then have to make a 180 degree turn before impinging upon the retinal rods, being transduced into an electrical signal in a most complicated process (Whikehart, 2003, pp.240-245) and, taking a tortuous path before finally ending up as ganglion cell spike trains proceeding along the optic nerves? It was almost as if nature was evolving the most efficient living macroscopic measurement apparatus possible, through which no superposed state could possibly be expected to escape wave function collapse, thereby possibly guaranteeing the survival of a species.

You can now begin to readily see that a superposed state or states would have an even more difficult if not impossible time running this gauntlet and remaining in a state of superposition as ganglion cell spike trains, since a condition of near simultaneity of passage of the branches of this superposed state would seem to have to be required from the very beginning, where the superposed state impinges upon the curved cornea and, be maintained to the very end, where the superposed state is supposed to collapse in the brain, mind or consciousness. This state of affairs would appear to be covered by Leggett (1984) who feels that "one might imagine that there are corrections to Schrödinger's equation which are totally negligible at the level of one, two or even one hundred particles but, play a major role when the number of particles involved becomes macroscopic". He is of course considering this problem from the standpoint of a collection of nonliving particles such as superconducting electrons, rather than a living macroscopic measuring device such as the eye. One can easily visualize by looking at Figs. 1-3, that the branches of any superposed state would also very rapidly begin to go out of phase at many different points in their transit of the eye, with possible implications for collapse, since the branches would no longer be simultaneous or in phase with regards to time and/or energy.

In a somewhat related vein Blokhintsev (Herbert, 1985) approaches this from the viewpoint of whether the process of wave function collapse occurs whenever a system's phases become sufficiently random, and shows that the process of *amplification*, which makes a quantum process visible to human eyes, will inevitably randomize quantum phases. It should be mentioned here that in the case of a living system such as the eye, there always has to be a preceding event of *transduction* before one can bring in the process of *amplification*

Shimony (1998) has addressed this issue of collapse in a very prescient fashion, with the conjecture that the locus of reduction is the macromolecules of the sensory and cognitive faculties, more specifically the photoreceptor protein of the rod cells of the eye, rhodopsin. One of the two components in rhodopsin is retinal, which can absorb a photon. In the resting state of retinal, hydrogen atoms attached to the eleventh and twelfth carbon atoms lie on the same side of the carbon backbone (so that the conformation is called *cis*), and this arrangement causes the backbone to bend. There is a potential barrier between the *cis* conformation and the *trans* conformation, in which the two hydrogen atoms mentioned are on opposite sides of the backbone from each other. But, when retinal in the *cis* conformation absorbs a photon, it acquires sufficient energy for a rotation to occur between the eleventh and twelfth carbon atoms, so that the *trans* conformation is achieved. Shimony's conjecture (1998) is that the reduction occurs at the retinal molecule itself: that there is a superselection rule operative which prevents a superposition of molecular conformations as different as *cis* and *trans* from occurring in nature.

When I posed the question to Markus Meister (2003) as to where the wave function might collapse or state reduction take place, he stated that, "My expectation is that the state vector collapses as soon as the photons cause a change in a classical system with lots of degrees of freedom. That would be the photoisomerization of rhodopsin in the retinal rods, which acts just like the blackening of a grain of film in the old two-slit experiment. Whether you take the film out of a Kodak canister or out of an eye should make little difference". His reasoning appears to be buttressed by experiments which demonstrate that the first step in vision, the *cis-trans* torsional isomerization of the rhodopsin chromophore (molecular shape-changing after the absorption of a photon), is essentially complete in only 200 femtoseconds (fs), which is one of the fastest photochemical reactions ever studied (Schoenlein et al, 1991; Baylor, 1996; Aalberts et al, 2000).

Matsuno (2003a) has commented upon the isomerization of rhodopsin from the standpoint of an internal measurement: "Robust transformation of a quantum such as a *cis-trans* transformation is an example of measurement internal to a molecule.

Measurement internal to a molecule is an activity of breaking a spacetime continuum on the part of the interacting electrons and atoms. The atoms provide the potential or the spacetime curvature to the moving electrons, and the electrons then exert the forces of push or pull upon the atoms. These two are not synchronous but sequential. If they are taken to be synchronous, the whole dynamic scheme develops in a unitary fashion as obeying the rule of linear quantum mechanics. In contrast, when these two movements of atoms and electrons are legitimately taken to be sequential, the atomic displacement following the electronic displacement, such as the destabilizing steric forces following the electronic excitation of a molecule (rhodopsin), is an indication of measurement proceeding internally. A consequence of such internal measurement is the appearance of a new discontinuity in the former spacetime continuum, like a *cis-trans* transformation".

Matsuno has also addressed another interesting area associated with the measurement problem, which deserves mention, as it has a bearing upon the issues being discussed in this paper. Some years ago he developed an approach (Matsuno, 2003b) saying that although the Schrödinger equation of the wave function is linear, the preparation of the boundary conditions is nonlinear because any material bodies are involved in internal measurement (i.e., nothing propagates faster than light). Locality upon the finiteness of light velocity may exert some nonlinear influence upon quantum nonlocality, causing entanglement and disentanglement. It may then be (Madrid, 2003) that boundary conditions imposed upon the Schrödinger equation determine the physical content of its solutions and, that vice versa, in order to obtain the solutions of the Schrödinger equation that describe a given physical situation, boundary conditions that fit the physical situation must be imposed upon the Schrödinger equation.

I asked Gary Shoemaker (2004) about the possibility of both continuous linear and discontinuous nonlinear wave function changes taking place within the eye and he stated that, "As I understand it the *key* is what happens to the Hamiltonian of the system during interaction. If the interaction changes the Hamiltonian of the quantum mechanical system (the one applicable during the formation of the quantum state) in such a way that the new Hamiltonian (original + interaction) has eigenstates different from the original eigenstates, then an irreversible measurement event will occur and this will force an irreversible change in the quantum state and 'break' any EPR-type 'connections'. My best guess is that it is highly likely that both scenarios will cause irreversible change".

There are at least two other processes which further guarantee that superposed states could never reach the brain, mind or consciousness:

1. There are some 2-3 action potentials or retinal ganglion cell (RGC) spike trains generated by the ganglion cell (Fig. 3) each time that a rhodopsin molecule is successfully activated and, which propagate along a single axon leading to the optic nerve (Barlow et al, 1971).

2. In addition, the information in the RGC spike trains is relayed to the visual cortex by lateral geniculate nucleus (LGN) relay cells (Fig. 1) operating in either of two regimes: tonic mode where each RGC spike is relayed by a single LGN spike or burst mode, where a single RGC input spike is relayed as a stereotyped burst of spikes (Reinagel et al, 1999).

The most interesting facet regarding item 2 is that it appears that the massive feedback pathway from the visual cortex to the LGN is involved in regulating the feed-forward properties of the LGN cells to selectively gate information transfer (Reinagel et al, 1999).

**3. Can *Euglena* collapse the wave function?**

I would now like to address the issue as to whether humans alone can collapse the wave function, since there is such a wide range of both vertebrates and invertebrates, all of whom preceded us on this planet, possessing varying degrees and types of vision processes from which our eye evolved (Wolken, 1967). As regards their degree of consciousness or if they can be considered sentient, I have attempted to deal with this in a very limited fashion in a previous paper (Thaheld, 2004) and this aspect lies outside the scope of the present paper. For the purpose of discussion I have chosen the *Euglena gracilis*, a unicellular protozoan (or more accurately an "algal flagellate") dating back around 2 billion years, approximately 50 μm long x 10 μm wide. It not only has the characteristics of a plant cell in the dark but, shares as well some of the attributes of an animal cell in the light (Wolken, 1967). As shown in Figs. 4 and 5 (Wolken, 1967) it possesses two different photoreceptors, an eyespot or stigma for light searching and, chloroplast for photosynthesis. It is believed that the eyespot directs the organism by phototrophic reactions to light of the right intensity and wavelength, to allow *Euglena* to carry on photosynthesis. You will also note in Fig. 5 the direct relationship between the photosensitive pigment of the eyespot and the flagellum as a sensory structure. This eyespot structure is analogous in many ways to the retinal rods of the eye and, it is interesting to note here that the number of photons which can excite the eyespot at the frequency of 500 nm is seven, while the human eye can detect a minimum of four photons, while individual rods can be activated by one photon (Rieke, Baylor, 1998).

The question now before us is: Are *Euglena* capable of collapsing the wave function of superposed states, either as an animal with its eyespot or as a plant with its chloroplast, in a manner analogous to the human eye? We will not address the matter of whether they can interpret the collapsed information or their degree of self-consciousness. We know that when photons with a wavelength of approximately 500 nm are incident upon the orange-red carotenoid pigment in the eyespot, they are converted into electrical signals which produces a movement of the flagellum known as *photomotion.* This is broken down into *photokinesis* or, the change in velocity or rate of swimming on illumination without regard to directed orientation and, *phototaxsis* the directed orientation of the organism to light of various wavelengths. The fact that we can communicate with the organism by means of the intensity and wavelength of light to the extent that its speed and direction of motion can be controlled suggests a sensory cell. The eyespot + flagellum may therefore be regarded as a servo- or feedback mechanism which maintains an optimal level of illumination for the organism.

Let us say that I have *Euglena* in a glass container placed several feet in front of me so that the superposed photon state from an experiment I am conducting impinges upon its eyespot before it impinges upon my eye. If the photons are transduced to an electrical pulse which results in the movement of the flagellum and the organism, would this not constitute bona fide evidence for the collapse of the wave function, just as real as if the superposed state had impinged upon my eye? I.e., it should make no difference to the superposed state what kind of eye or visual apparatus it impinges upon, the only criteria is that it be a living entity, which covers a wide range. Based on this analysis, this would mean that only measured information would now reach my eye and brain since the *Euglena* would have already collapsed the superposed state on its own as a result of transduction of the superposed photons to an electrical pulse and, resulting movement of the flagellum and the organism directly tied to this photon input.

Now, let us examine if the chloroplast might possess the same ability as the eyespot to collapse the wave function of a superposed state. As regards the chloroplasts, their main photosynthetic pigment is comprised of the green pigments or chlorophylls which are sensitive to light in the 600 nm range. They vary in number from one to more than 20 in each *Euglena* depending upon the different species and varieties. Photons in the visible photosynthetic part of the spectrum have sufficient energy to bring about, when they are absorbed, transitions from one electronic energy level (usually the ground state) to another (Kirk, 1983). Within a complex molecule such as chlorophyll or, any of the other photosynthetic pigments, there is usually more than one possible electronic energy transition which can occur.

Regarding the absorption process leading from a photon to photosynthesis, the energy of a molecule can be considered to be part rotational, part vibrational and part electronic. A molecule can only have one of a discrete series of energy values. Energy increments corresponding to changes in a molecule's electronic energy are large, those corresponding to changes in vibrational energy are intermediate in size and, those corresponding to changes in rotational energy are small. Molecules can obtain energy from radiation as well as from other molecules. When a photon passes within the vicinity of a molecule, there is a finite probability that it will be captured by that molecule and be absorbed, with the energy of the molecule being increased by an amount corresponding to the energy of the photon. I.e., the capture of a photon by a molecule results in the simultaneous transition of an electron in that molecule from the ground state to an excited state. Within a complex molecule such as chlorophyll, there is usually more than one possible electronic transition which can occur. Any given electronic transition is preferentially excited by light which has an amount of energy per photon corresponding to the energy required for the transition. Absorption of a blue photon leads to a substantially higher energy level than absorption of a red photon. Immediately after absorption of a blue photon there is a very rapid series of transitions downwards through the various rotational/vibrational levels (associated with transfer of small increments of rotational/vibrational energy to adjoining molecules) until the lower electronic energy state is reached. It is the energy in this lowest excited singlet state which is used in photosynthesis and, it is because an excited chlorophyll molecule usually ends up in this state anyway, that all *absorbed* visual quanta are equivalent. Thus, most of the light energy absorbed by the chlorophyll molecules ends up as chemical energy in the form of photosynthetically produced biomass (Kirk, 1983).

You can now readily see that if a superposed photon state impinges upon a chloroplast, that the photon energy is imparted to an electron as the first step in the beginning of photosynthesis. And, that it is here that the superposed state probably collapses when the energy level of the chlorophyll molecule increases when a photon interacts with a molecule or, when an electron is transferred from a donor molecule to an acceptor molecule. We are not interested at this stage in a complete description of a very complicated photosynthetic process, only the initial transfer of the energy from an incident photon to a molecule and thence to an electron, which should serve to reduce the state vector (Kirk, 1983; Leedale, 1967). It is very hard in this scenario to attempt to visualize a superposed photon state interacting with a chlorophyll molecule, and assuming that there is a transfer of the superposed state to an electron rather than the immediate collapse of this state. This means that the chloroplast can collapse superposed photon states just like the eyespot can, except in a different fashion.

**4. Proposed Ghirardi superposed photon-retinal tissue experiment**

It is anticipated that sometime in the near future an experiment will be conducted at the Univ. of Milan utilizing retinal tissue mounted on a microelectrode array and, superposed single photon states (Thaheld, 2000; 2003) to test Ghirardi's theory (Ghirardi, 1999), as to whether superposed states continue on past the retina and are collapsed in the visual cortex of an observer, due to what is called a Spontaneous Localization process (Ghirardi et al, 1986; Aicardi et al, 1991). This retinal tissue will have its outer elements such as the sclera, pigment epithelium and the inner limiting membrane removed (Meister et al, 1994). Since these usually account for large photon transmission losses, this removal of these elements will allow us to direct superposed single photons against the retinal rods with no intervening losses of any of the photons and, subject only to the normal quantum detection efficiencies of the rods, which range from 25% to 36%.

It should be pointed out here that there appears to be one problem with his theory. He states that, "As soon as the superposition of the two stimuli excite the retina, two nervous signals start and propagate along two different axons, still in a state of superposition". As has already been pointed out in Sec. 2 and demonstrated by prior retinal research, it is physically impossible for two superposed nervous signals to start and propagate along two different axons. This same analysis also applies to a recent paper (Schlosshauer, 2005), in which a variation of the Ghirardi theory is proposed in which the photon-rhodopsin interaction leads to a superposition of the *cis-trans* states of the rhodopsin molecule, which states are further correlated with neuronal states in the visual cortex, where decoherence will supposedly lead to a single subjective outcome.

The results of this experiment should tell us whether collapse of the wave function takes place in the brain, implying action by the mind or consciousness or, if it takes place much sooner within the retinal apparatus of the eye, lending credibility to the theory advanced in this paper, in agreement with and confirming the position taken by Shimony and Wigner.

## 5. The "many worlds" interpretation of the measurement problem: Can biophysics tell us something about quantum cosmology?

There is another approach to dealing with the measurement problem, popularly known as the "*many worlds interpretation*". According to Everett (1957; Shimony, 1963) the measurement problem can be solved at one stroke by simply assuming that *no collapse or state reduction ever takes place.* The whole issue of the transition from 'possible' to 'actual' is taken care of in his theory in a very simple way, *there is no such transition.* If there is no collapse this means that we do not need von Neumann's two distinct types of quantum mechanical process; we do not use the idea of measurement selecting a particular branch or component of the evolving wave function; we have the universal wave function of Everett (Whitaker, 1996. pp. 274-284) representing *both* possibilities i.e., the various branches of the wave function co-exist.

While Everett avoids many of von Neumann's problems, he does so at the expense of failing drastically to ensure the aim of von Neumann's procedure that, following a measurement, the apparatus is left in a distinct state and, in particular, that the mind of the observer recognizes a distinct result for the measurement.

Expanding upon and popularizing Everett's theory, de Witt (1973) proposes that at the moment of measurement there is a splitting into two different worlds, one for each distinct measurement, with further splitting at every subsequent measurement, until one arrives at a *many worlds* or *many universes interpretation.* In this interpretation of quantum cosmology all the amplitudes of a given spatial section and field configurations in the universe, are pieced together to give amplitudes for cosmological histories which are all equally real and co-exist in superposition (Everett, 1957; Aguirre, Tegmark, 2004).

Wigner (Mehra, 1995. pp. 68, 112; Esfeld, 1999) refuses to dissolve the measurement problem by countenancing a split of the world instead of a state reduction. Mentioning the *many worlds interpretation*, he dismisses the notion of a state function of the universe as senseless. Bell (1987) has also raised serious objections to Everett's theory.

If the experiment proposed in previous papers (Thaheld, 2000; 2003) is successful, this may finally reveal that definite and irreversible wave function collapse results from measurements performed on superposed states by a living macroscopic detector like the eye, in a discontinuous nonlinear fashion. This would appear to rule out the *many worlds interpretation*. Thus it is that biophysics, in the role of the eye as a non-conscious, living, macroscopic measuring apparatus, may have a role to play in the resolution of certain theories regarding quantum cosmology and the measurement problem.

**6. Discussion**

If the theory advanced in this paper is correct, we will have been successful in moving the previous site for wave function collapse out of the brain and mind or consciousness (what is known as the subjective), some 17 cm *back* from the visual cortex to the eye. It is at this point that we can consider two different types of objective wave function changes to take place in a sequential fashion, continuous and linear in the first phase and discontinuous and nonlinear in the second phase. The continuous linear phase would encompass the area of the eye between the cornea and the retinal photoreceptors, while the discontinuous nonlinear phase involves the rhodopsin molecules in the photoreceptors.

We know that normally approximately 90% of non-superposed photons are lost in this first phase, including those that do not successfully impinge upon and activate rhodopsin molecules in the photoreceptors. Whether these losses will increase or decrease as a result of the photons being in superposed states is not known at this time, and so it is that we have to make certain assumptions. For the purposes of our discussion it makes no difference whether the superposed photon losses are higher or lower than the normal 90%, as this would be adequately compensated for in the second phase of this process through 100% collapse of the wave function of any remaining superposed photons.

So, for the sake of convenience, let us assume that roughly 90% of the superposed photons initially incident upon the cornea will be subjected to a *continuous change of state* as they traverse the medium consisting of the aqueous and vitreous humors and, portions of the retina including the sclera, pigment epithelium and the inner limiting membrane, a length of about 2 cm. These changes will appear at the macro-level to be a classical deterministic change (i.e., can be described adequately with classical dynamical equations) but ultimately at the micro or quantum level, the photon is being forced into new (albeit fairly close), neighboring quantum states as it travels, that is, the photon experiences a large number of closely spaced events that with each event only slightly, but irreversibly, changes its state (Shoemaker, 2004).

Let us now consider the second phase of this process, the *discontinuous change of state.* This would commence with a successful impingement of one branch of the superposed photon state upon a rhodopsin molecule of about 38 kDa, containing a retinal chromophore consisting of some 40 atoms, which in turn has an active site of about 10 Angstroms. The structural shape of this active site is changed in 200 fs after the absorption of a photon, in what is known as the *cis-trans* torsional isomerization of the rhodopsin chromophore site (Mathies, 2003; Schoenlein et al, 1991).

The way to think about it is to consider the angular rotation about the 11=12 bond. This bond rotates to about 90 degrees in the first 100 fs (with initial significant motions occurring in about 50 fs) and then goes to a formally *trans* configuration about the 11=12 bond in 200 fs (Mathies, 2003). The molecular shape change is the classical signal of the photoabsorption quantum event. In addition, the vibrational spectrum also evolves as the molecule is changing its shape (Schoenlein, 1991). It is felt by the author that somewhere within the area of this initial activation or rotation of the bond that the *discontinuous change of state* or the measurement might occur, perhaps even prior to the *amplification* process.

It would now begin to appear that at this objective level we might be getting very close to what is known as the elusive Heisenberg 'cut', which divides the quantum and classical world, where possibility changes into actuality and linear becomes nonlinear (Herbert, 1985).

It would appear that either in the *Euglena* eyespot-flagellum interface or, in the initial stage of the photosynthesis process of the chloroplast, that we might also be looking at the equivalent area where the 'cut' could take place. This step could be achieved when the photon raises an electron from the ground state to one of two possible excited singlet states depending upon the wavelength and energy of the photon, with the result that after absorption of a photon there is a very rapid series of transitions downwards through the various rotational/vibrational levels (associated with transfer of small increments of rotational/vibrational energy to adjacent molecules) until the lower electronic state (ground state) is reached (Kirk, 1983).

If it can be demonstrated that *Euglena* can collapse the wave function either through its eyespot-flagellum combination as an animal or, through its chloroplasts as a plant, since both of these processes are very simple and basic, this would tell us several very important things. First, that it may then be possible for any living entity, including the most primitive and simple plant or animal, vertebrate or invertebrate, to collapse the wave function and, that the first life on earth or in the universe was able to perform this process in a discontinuous nonlinear fashion. Second, that we may have to expand the use of the terms 'sentient' and 'consciousness' to include these primitive life forms, as we will be observing already measured information resulting from their ability to collapse superposed states.

## 7. Conclusion

It is interesting to note that von Neumann, London and Bauer, and Wigner were able to come up with such definite pronouncements on this subject of measurement, which is built upon what appears to be a shaky foundation of brain, mind and consciousness. One is also struck by the fact that none of these mathematicians or physicists had any real grounding in biology or psychology and yet, when they encountered this problem, they turned to the brain, mind and consciousness in an attempt to arrive at a solution. They appear to have been correct in one respect, and that is that measured information finally does get received by the brain and enters into consciousness and the mind.

In retrospect it becomes most fascinating to once again review some of the issues which were involved in an attempt to see how and why the measurement problem ever arose (Herbert, 1985). It all started when von Neumann postulated that the possible states of a system are characterized by state vectors or wave functions that can change in two ways, continuously and discontinuously. This discontinuous change, reduction of the state vector or collapse of the wave function, raised the question as to how and when does the wave function collapse or state reduction occur. von Neumann's all-quantum description will not work unless such a collapse really occurs as a physical process in every quantum measurement (Herbert, 1985). von Neumann was anxious to find a natural location for the wave function collapse, which is essential for his interpretation of quantum theory and, visualized the measurement act as broken into small steps called the von Neumann chain.

While searching for a place to break this chain he showed that one can cut this chain and insert a collapse anywhere you want. Attempting to separate the world unmeasured and measured resulted in a logic gap, so von Neumann was forced to seize upon the only peculiar link in the chain, the process by which the physical signal in the brain becomes an experience in the human mind and thus, concluded that human consciousness is the site of the wave function collapse (Herbert, 1985).

It is to be hoped that the theory proposed in this paper will help to finally begin to lay the measurement problem to rest after these many decades of controversy, thereby enabling us to address certain unresolved issues, with emphasis on the following:

1. That the brain, mind or consciousness play no subjective role in the collapse of the wave function, with this event taking place naturally in an objective and stochastic discontinuous nonlinear fashion within the complex architecture of the eye. This means that only non-superposed states or final, measured information reaches the brain, mind or consciousness.

2. That while the macroscopic measuring instrument known as the eye is an integral part of the brain, it is a non-conscious entity, such that the brain, mind or consciousness can have no subjective effect upon its objective and stochastic wave function collapse processes.

3. That wave function collapse is a *real physical process* of a discontinuous nonlinear nature, resulting when a superposed microscopic system interacts with a living macroscopic measuring instrument, in this instance the eye.

4. That it may now be possible to pinpoint the specific location of wave function collapse, in the case of the eye, to a region of the retinal chromophore known as the active site, involving an area of less than 10 Angstroms and a time scale measuring a few fs. This would also get one very close to the division between the quantum and classical worlds or the Heisenberg 'cut' and, opens up heretofore unattainable experimental possibilities.

5. If wave function collapse is a *real physical process* with definite *irreversible* outcomes, this means that the *many worlds* or *many universes interpretation* of the measurement problem is probably incorrect.

6. If the measurement problem is resolved in the fashion outlined in this paper, this would require that the Schrödinger linear equation has to be modified to include nonlinear discontinuous changes. Furthermore, since the quantum definition of physical units (such as the second, the meter, the Ohm or electrical resistance, and the electrical charge, among others) heavily rely on the validity of basic dynamical equations like the Schrödinger equation, if this was to be modified, then some definitions of physical units may also be affected (Amelino et al, 2005).

7. This would also mean that Schrödinger's cat (Schrödinger, 1935) can never be in a superposed state of dead and alive for more than a fleeting instant, if at all, due to its constant interaction with the decohering environment via its visual processes, among other things. This would arise as a result of a constant massive stream of non-superposed photons being converted into electrical pulses which impinge upon the brain. What it means additionally however, is that although living macroscopic superposed states are collapsed instantly, living macroscopic *entangled states* can resist collapse and regenerate or maintain their entangled status for long periods of time after repeated measurements, as has already been revealed in several experiments (Pizzi et al, 2004; Standish et al, 2004; Thaheld, 2004; 2005; Wackermann et al, 2003).

8. That any living system, whether animal or plant, possesses this same ability to collapse the wave function either through its visual processes or through chlorophyll or chloroplast structures. And, that while vision has been stressed in this paper, it can serve as a model for all the other senses.

**Acknowledgement**

I wish to thank the reviewers and the editor for their valuable comments and suggestions, and also Denis Baylor, Andrei Khrennikov, Richard Mathies, Koichiro Matsuno, Markus Meister, Fred Rieke and especially Gary Shoemaker for answering my many questions. Also, Michael Esfeld whose Essay Review of Wigner's View of Physical Reality provided a wealth of information on Wigner's thinking regarding the measurement problem. And, once again, to Thesa von Hohenastenberg-Wigandt for first getting me started in this field of endeavor and for her advice and encouragement.

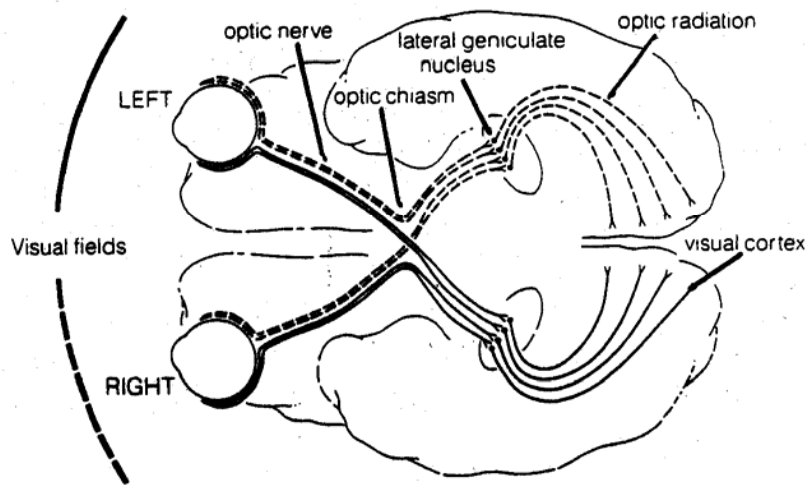

Fig. 1

Fig. 1 – Diagram of the visual pathways in primates viewed from the underside of the brain (reprinted with permission from Harvard University Press. John Dowling, 1987).

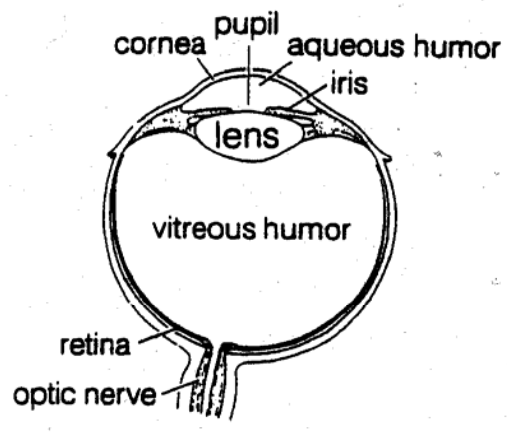

Fig. 2

Fig. 2 – Schematic drawing of a primate eye (reprinted with permission from Harvard University Press. John Dowling 1987).

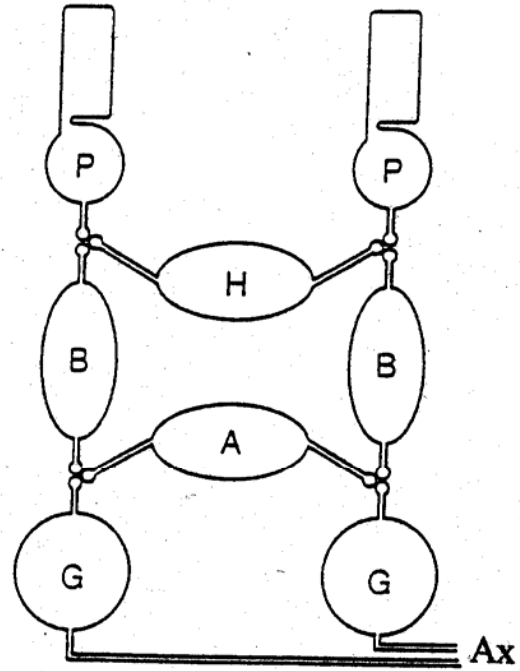

Fig. 3

Fig. 3 – Schematic diagram of the vertebrate retina showing photoreceptors (P), horizontal cells (H), bipolar cells (B), amacrine cells (A) and ganglion cells (G) with their axons (Ax) leading to the optic nerve. (Reprinted from Meister et al. (1994), Copyright 1994, with permission from Elsevier Science).

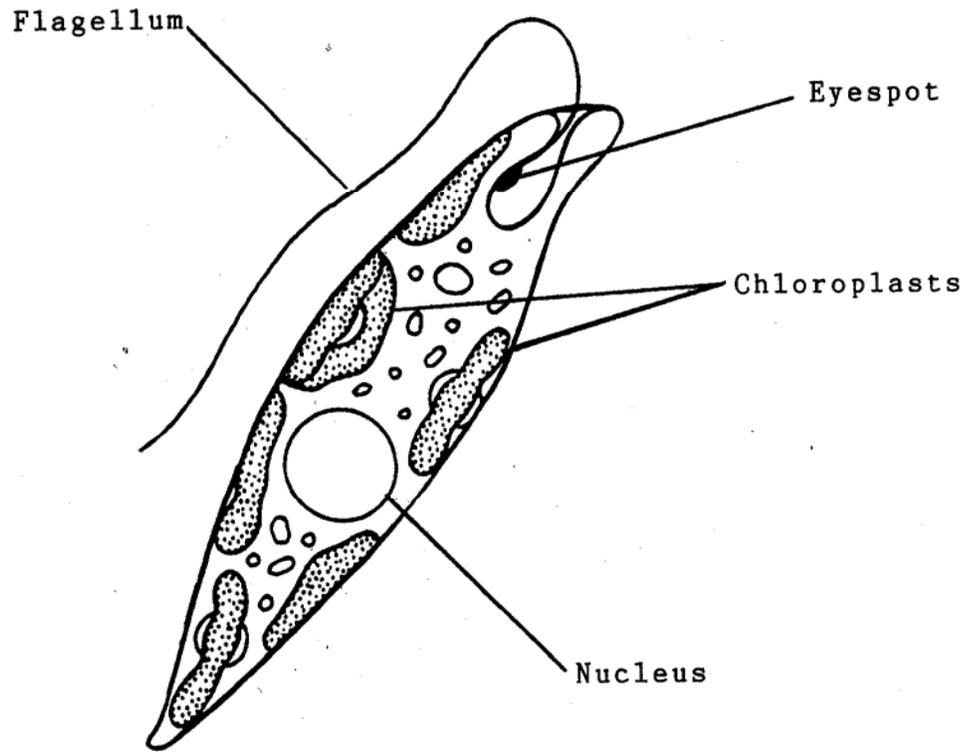

Fig. 4 – *Euglena gracilis* (light grown) (reprinted with permission from Meredith Publishing Co. J. Wolken, 1967).

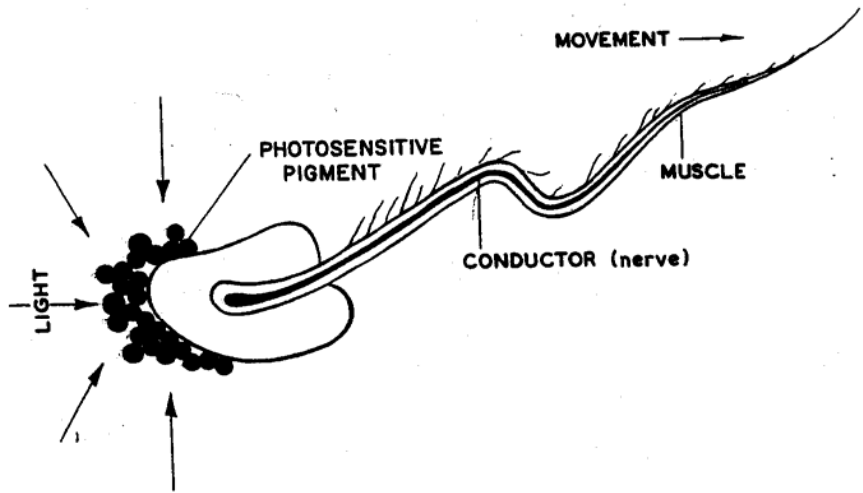

Fig. 5 – Detail of the eyespot and flagellum of *Euglena gracilis.* (Reprinted with permission from Meredith Publishing Co. J. Wolken, 1967).